# One approach for determining susceptibilities and order parameters in multi-band Hubbard model

**Sung-Jin Oh, Hak-Chol Pak, and Kuk-Chol Ri**

**Department of Physics, Kim Il Sung University, DaeSong District, Pyongyang, DPR Korea**

## Abstract

We present an approach for determining susceptibilities and order parameters in multi-band Hubbard model within functional renormalization group method, and apply it to study various instabilities of the FeAs-based high temperature superconductor. First, we derive the formulae of susceptibilities and order parameters of superconducting parings, spin density waves and charge density waves with diverse wave vectors which could occur in multi-band Hubbard model. Second, we apply it to the FeAs-based high temperature superconductor and find an electronic-driven superconducting pairing instability within a five band model with pure repulsive interactions. Our study shows that for doping of our concern, there is competition between antiferromagnetic and superconducting instabilities. In addition, we show that for doping of 0.1, the susceptibility of extended *s*-wave pairing is dominant over others, while a staggered superconducting pairing with $\mathbf{Q} = (\pi, 0)$ has the second largest value of susceptibility.

## 1. Introduction

Recently discovered FeAs-based high temperature superconductors $MFeAsO_{1-x}F_x$[1] have a phase diagram similar to the copper oxide superconductors[2] and their superconducting transition temperature can be raised to 55K by varying the rare earth element M included in them[3]. Therefore, in recent years, these materials have attracted intense investigation in the field of condensed matter physics. In the structure of $MFeAsO_{1-x}F_x$, As-Fe-As tri-layers are separated by $MO_{1-x}F_x$ spacers[4].

The FeAs-based high temperature superconductors (the iron pnictides) have some common features with cuprates. Both of them have layered structures with separated conducting layers and charge reservoir and can be considered as quasi-two dimentional systems. The ground states of their stoichiometric parent compounds have antiferromagnetic (AFM) order. Neutron scattering analysis shows that the parent compound of the iron pnictides has stripe AFM order with the wave vector $\mathbf{Q} = (0, \pi)$ [5, 6]. They have very similar phase diagrams. Especially, similar to the cuprates, the AFM order is quickly replaced by superconductivity with increasing doping, suggesting a deep relation between antiferromagnetism and superconductivity in the iron pnicides[7, 17].

But there are some differences between the iron pnictides and the cuprates. Unlike the cuprates, the stoichiometric parent compounds of the iron pnictides are metallic (or semimetallic) antiferromagnets[5, 9] rather than antiferromagnetic Mott insulators. There is strong hybridization of Fe and As orbitals and there are several bands sharing Fermi surfaces in the iron pnictides[4] while the only one band contains Fermi surface in the cuprates. So the FeAs-based high temperature superconductors should be modeled by multi-band Hubbard model. The electron-electron correlation of the iron pnictides is weaker than that of cuprates[4, 8]. This fact causes us to select the functional renormalization group (FRG) method to calculate the electronic property.

These common and different properties suggest some relation between the superconducting mechanism of both superconductors. Now, it is expected that progresses in understanding this new class of high temperature superconductor will also help to solve the old problem – the superconducting mechanism in the cuprates.

According to the electronic structure calculations[10] and the angle-integrated photoemission spectroscopy[11], the carriers at Fermi energy are mainly induced from Fe atomic orbitals. Hence, we will focus on a As-Fe-As tri-layer.

Each unit cell in the tri-layer, in principle, contains two Fe atoms from the view of its translational invariance, but we can take a unit cell containing only one Fe atom instead, considering that it has the plane inversion symmetry accompanied by fractional translation. Thus, we can model it as a multi-band Hubbard lattice consisting of five Fe *d*-orbitals.

In this paper we present an approach for determining various kinds of susceptibilities and order parameters. It is based on the patch approximation of the FRG method in which all k-points in one patch of the Brillouin zone (BZ) are represented by one point on the Fermi surface.

Then we apply it to $LaFeAsO_{1-x}F_x$ to study diverse instabilities and determine the pairing symmetry for this material.

## 2. The various kinds of susceptibilities and order parameters in multi-band Hubbard model

It is very important in study of phase transition to determine a type of instability and a symmetry of order parameters. Usually, in the FRG method[12], one considers the possible instabilities by considering divergence of susceptibilities at a critical point starting from a disordered phase. In detail, we bring in diverse additional Hamiltonians of superconducting pairing, ferromagnetic or anti-ferromagnetic ordering etc., respectively, and calculate the corresponding susceptibilities using renormalized vertex functions obtained in FRG method.

The type of strongest divergent instability would be selected by comparing the divergences of various susceptibilities. We can expect the phase transition of this type if its divergent susceptibility exceeds certain critical value. We develop an approach for determining the type of instability and the shape of order parameters in multi-band Hubbard model following this procedure.

For a singlet paring, the additional Hamiltonian is

$$\hat{H}_{ext} = -\frac{\lambda}{2}[\sum_{m,m',\mathbf{k},\sigma} S^*(m,m',\mathbf{k})\sigma \hat{c}^+_{m;\mathbf{k},\sigma}\hat{c}^+_{m';\mathbf{Q}-\mathbf{k},-\sigma} + \sum_{m,m',\mathbf{k},\sigma} S(m,m',\mathbf{k})\sigma \hat{c}_{m';\mathbf{Q}-\mathbf{k},-\sigma}\hat{c}_{m;\mathbf{k},\sigma}] \quad (1)$$

where $\sigma = +1(\uparrow), -1(\downarrow)$ is spin polarization, $S(m,m',\mathbf{k}) = S(m',m,\mathbf{Q}-\mathbf{k})$ is coupling field, and $\mathbf{Q}=(0,0)$, $\mathbf{Q}=(\pi,\pi)$, or $\mathbf{Q}=(\pi,0)$. Corresponding action is given by

$$S_{ext} = -\frac{\lambda}{2}\sum_{m,m';\mathbf{k},\omega_n,\sigma}\frac{\sigma}{\hbar}S^*(m,m',\mathbf{k})\overline{\psi}_{m;\mathbf{k},\omega_n,\sigma}\overline{\psi}_{m';\mathbf{Q}-\mathbf{k},-\omega_n,-\sigma}$$
$$-\frac{\lambda}{2}\sum_{m,m';\mathbf{k},\omega_n,\sigma}\frac{\sigma}{\hbar}S(m,m',\mathbf{k})\psi_{m';\mathbf{Q}-\mathbf{k},-\omega_n,-\sigma}\psi_{m;\mathbf{k},\omega_n,\sigma} \quad (2)$$

The free energy is constructed using this action.

$$\Omega = \Omega_0 - \frac{1}{\beta}\ln(W/Z_0),$$
$$W \equiv \int D\overline{\psi}D\psi \exp\{(\overline{\psi},[g^0]^{-1}\psi) - S_{int}(\{\overline{\psi}\},\{\psi\}) - S_{ext}(\{\overline{\psi}\},\{\psi\};\lambda)\}$$

where $\Omega_0$, $Z_0$ is the free energy and the partition function of the system of free particles.

The susceptibility is obtained by differentiating twice the free energy with respect to $\lambda$.

$$\chi_{\text{singletSC}} = -\frac{\partial^2 \Omega}{\partial \lambda^2}\bigg|_{\lambda=0} = \frac{1}{\beta}\frac{\partial}{\partial \lambda}\left(\frac{1}{W}\frac{\partial W}{\partial \lambda}\right)\bigg|_{\lambda=0} = \frac{1}{\beta}\left(\frac{1}{W}\frac{\partial^2 W}{\partial \lambda^2}\right)\bigg|_{\lambda=0} - \frac{1}{\beta}\left(\frac{1}{W}\frac{\partial W}{\partial \lambda}\right)\bigg|_{\lambda=0} \cdot \left(\frac{1}{W}\frac{\partial W}{\partial \lambda}\right)\bigg|_{\lambda=0}$$

After some calculations, this can be written as

$$\chi_{\text{singletSC}} = \frac{1}{2\beta\hbar^2}\sum_{m,m';\mathbf{k},\omega_n,\sigma}\sum_{n,n';\mathbf{k}',\omega'_n,\sigma'}S^*(n,n',\mathbf{k}')S(m,m',\mathbf{k})\sigma\sigma'$$
$$\times \left\langle \psi_{m;\mathbf{k},\omega_n,\sigma}\psi_{m';\mathbf{Q}-\mathbf{k},-\omega_n,-\sigma}\overline{\psi}_{n';\mathbf{Q}-\mathbf{k}',-\omega'_n,-\sigma'}\overline{\psi}_{n;\mathbf{k}',\omega'_n,\sigma'}\right\rangle_{\lambda=0}$$

Using the relation between 4-point Green's function and vertex function and taking an approximation of $g(m,\mathbf{k},\omega_n) \approx g^0(m,\mathbf{k},\omega_n)$, we obtain the following result.

$$\frac{1}{N}\chi_{\text{singletSC}} = \frac{2}{N}\sum_{m,m',\mathbf{k}}|S(m,m',\mathbf{k})|^2 L(m,m',\mathbf{k})$$
$$-\frac{2}{N}\sum_{m,m',\mathbf{k}}\frac{1}{N}\sum_{n,n',\mathbf{k}'}S^*(m,m',\mathbf{k})L(m,m',\mathbf{k})S(n,n',\mathbf{k}')L(n,n',\mathbf{k}')$$
$$\times \frac{1}{2}[V(n,\mathbf{k}';n',\mathbf{Q}-\mathbf{k}';;m,\mathbf{k};m',\mathbf{Q}-\mathbf{k}) + V(n',\mathbf{Q}-\mathbf{k}';n,\mathbf{k}';;m,\mathbf{k};m',\mathbf{Q}-\mathbf{k})] \quad (3)$$

where $L(m,m',\mathbf{k}) \equiv \frac{1}{\beta\hbar^2}\sum_{\omega_n}g^0(m,\mathbf{k},\omega_n)g^0(m',\mathbf{Q}-\mathbf{k},-\omega_n) = L(m',m,\mathbf{Q}-\mathbf{k})$.

Value of susceptibility may be different according to a concrete form of $S(m,m',\mathbf{k})$. In this case we can regard $S_{\max}(m,m',\mathbf{k})$ corresponding to maximum of susceptibility as a candidate of actual superconducting gap function. In other words, it could be expected that the gap function is characterized in its form by $S_{\max}(m,m',\mathbf{k})$ when pairing really occurs under given conditions. But it is necessary to give a normalization condition in order to determine $S_{\max}(m,m',\mathbf{k})$. We can take the normalization condition such that a susceptibility of the system of independent particles

$$\frac{2}{N}\sum_{m,m',\mathbf{k}}|S(m,m',\mathbf{k})|^2 L(m,m',\mathbf{k})$$

is a constant. Our normalization condition is

$$\frac{2}{N}\sum_{m,m',\mathbf{k}}|S(m,m',\mathbf{k})|^2 L(m,m',\mathbf{k}) = 1\text{eV} \tag{4}$$

Under this condition, the susceptibility and the gap function is given by a largest eigenvalue and eigenvector of equation:

$$-\frac{1}{N}\sum_{n,n',\mathbf{k}'}\frac{1}{2}[V(n,\mathbf{k}';n',\mathbf{Q}-\mathbf{k}';;m,\mathbf{k};m',\mathbf{Q}-\mathbf{k})+V(n',\mathbf{Q}-\mathbf{k}';n,\mathbf{k}';;m,\mathbf{k};m',\mathbf{Q}-\mathbf{k})]$$

$$\times L(n,n',\mathbf{k}') S(n,n',\mathbf{k}') = \chi' \cdot S(m,m',\mathbf{k}) \tag{5}$$

$$\frac{1}{N}\chi_{\text{singletSC}} = (1+\chi'_{\max})\text{eV} \tag{6}$$

In patch approximation[13], all the points of momentum space within one patch is represented by a point $\mathbf{k}_{Fi}$ on Fermi surface belonging to it and the vertex functions in this region are considered as equal with each other.

$$V(m_1,\mathbf{k}_1;m_2,\mathbf{k}_2;;m_3,\mathbf{k}_3;m_4,\mathbf{k}_4) \approx V(m_1,\mathbf{k}_{F1};m_2,\mathbf{k}_{F2};;m_3,\mathbf{k}_{F3};m_4,\mathbf{k}_{F4}) \tag{7}$$

As you see from (5), the coupling constant $S(m,m',\mathbf{k})$ at $\mathbf{k}$ point belonging to m-th band, i-th patch has the form of

$$S(m,m',\mathbf{k}) = S(m,m',\mathbf{k}_{Fi}) \qquad \{\mathbf{k} \in \text{patch } i \text{ of band } m\} \tag{8}$$

Thus, equation (5) is equivalent to

$$\sum_{n,n',j} T(m,m',i;n,n',j) X(n,n',j) = \chi' \cdot X(m,m',i), \qquad \sum_{m,m',i}|X(m,m',i)|^2 = 1\text{eV} \tag{9}$$

with

$$L(m,m',i) \equiv \frac{1}{N}\sum_{\mathbf{k}}^{i} L(m,m',\mathbf{k}), \quad X(m,m',i) \equiv \sqrt{2L(m,m',i)} S(m,m',\mathbf{k}_{Fi}), \tag{10}$$

$$T(m,m',i;n,n',j) \equiv -\frac{1}{2}\sqrt{L(m,m',i)}\sqrt{L(n,n',j)}[V(n,\mathbf{k}_{Fj};n',\mathbf{Q}-\mathbf{k}_{Fj};;m,\mathbf{k}_{Fi};m',\mathbf{Q}-\mathbf{k}_{Fi})$$

$$+V(n',\mathbf{Q}-\mathbf{k}_{Fj};n,\mathbf{k}_{Fj};;m,\mathbf{k}_{Fi};m',\mathbf{Q}-\mathbf{k}_{Fi})] \tag{11}$$

Order parameter is defined as and proportional to

$$\Delta_{\text{singletSC}}(m,m';\mathbf{k}) = \sum_{\sigma}\sigma\langle\hat{c}_{m';\mathbf{Q}-\mathbf{k},-\sigma}\hat{c}_{m;\mathbf{k},\sigma}\rangle \propto S^*_{\max}(m,m',\mathbf{k}_{Fi})L(m,m',\mathbf{k})$$

$$\{\mathbf{k} \in \text{patch } i \text{ of band } m\} \tag{12}$$

In the same way, the susceptibilities of triplet pairing, spin density wave (SDW) and charge density wave (CDW) are obtained. All the results are represented in the form of (9), except that the expression for $T(m,m',i;n,n',j)$ is different from each other.

Triplet pairing : $\quad T(m,m',i;n,n',j) \equiv -\frac{1}{2}\sqrt{L(m,m',i)}\sqrt{L(n,n',j)}$

$$\times [V(n,\mathbf{k}_{Fj};n',\mathbf{Q}-\mathbf{k}_{Fj};;m,\mathbf{k}_{Fi};m',\mathbf{Q}-\mathbf{k}_{Fi}) - V(n',\mathbf{Q}-\mathbf{k}_{Fj};n,\mathbf{k}_{Fj};;m,\mathbf{k}_{Fi};m',\mathbf{Q}-\mathbf{k}_{Fi})] \tag{13}$$

Spin density wave : $\quad T(m,m',i;n,n',j) \equiv \sqrt{L(m,m',i)}\sqrt{L(n,n',j)}$

$$\times V(m,\mathbf{k}_{Fi};n',\mathbf{k}_{Fj}+\mathbf{Q};;n,\mathbf{k}_{Fj};m',\mathbf{k}_{Fi}+\mathbf{Q}) \tag{14}$$

Charge density wave : $\quad T(m,m',i;n,n',j) \equiv \sqrt{L(m,m',i)}\sqrt{L(n,n',j)}$

$$\times [V(m,\mathbf{k}_{Fi};n',\mathbf{k}_{Fj}+\mathbf{Q};;n,\mathbf{k}_{Fj};m',\mathbf{k}_{Fi}+\mathbf{Q}) - 2V(m,\mathbf{k}_{Fi};n',\mathbf{k}_{Fj}+\mathbf{Q};;m',\mathbf{k}_{Fi}+\mathbf{Q};n,\mathbf{k}_{Fj})] \tag{15}$$

For the SDW and CDW, $L(m,m',i)$ is

$$L(m,m',i) \equiv \frac{1}{N}\sum_{\mathbf{k}}^{i} L(m,m',\mathbf{k}),$$

$$L(m,m',\mathbf{k}) \equiv -\frac{1}{\beta\hbar^2}\sum_{\omega_n} g^0(m,\mathbf{k},\omega_n)g^0(m',\mathbf{k}+\mathbf{Q},\omega_n) = L(m',m,\mathbf{k}+\mathbf{Q}).$$

## 3. Various instabilities in FeAs-based high temperature superconductors

In previous approaches for determining susceptibilities and order parameters, one has to set initial values of order parameters and integrate their renormalization group equations simultaneously with integrating that of vertex function. It is necessary in study of diverse instabilities to postulate all possibility of their initial values, which makes trouble, especially, for the multi-band system.

But our approach has an advantage that it does not need to set initial values of order parameters and integrate them. It needs only the final FRG result of vertex function.

Now, we apply our approach to $LaFeAsO_{1-x}F_x$ to study diverse instabilities and determine the pairing symmetry for this material. It is very important in study of pairing mechanism and transport properties to determine the symmetry of the superconducting gap function.

There are many results of experimental and theoretical study on the symmetry of the gap function. In literature [14], they calculated the vertex function using FRG method and applied the mean field approximation to find the gap function. Such a choice is not thought to be reasonable. We calculate vertex function $V(m_1,\mathbf{k}_{F1};m_2,\mathbf{k}_{F2};;m_3,\mathbf{k}_{F3};m_4,\mathbf{k}_{F4})$ of the material $LaFeAsO_{1-x}F_x$ in the same way and consider possible instabilities using the approach described above.

In this paper we study a five band model with Hubbard-like and Hund interactions which can be written as

$$\hat{H} = \sum_{\mathbf{k},\sigma} \sum_{\alpha,\beta=1}^{5} \hat{c}^+_{\mathbf{k}\alpha\sigma} K_{\alpha\beta}(\mathbf{k}) \hat{c}_{\mathbf{k}\beta\sigma} + \hat{H}_I \quad (16)$$

$$\hat{H}_I = \sum_i \{U_1 \sum_\alpha \hat{n}_{i\alpha\uparrow}\hat{n}_{i\alpha\downarrow} + U_2 \sum_{\alpha<\beta} \hat{n}_{i\alpha}\hat{n}_{i\beta} +$$

$$+ J_H [\sum_{\alpha<\beta} \sum_{\sigma,\sigma'} \hat{c}^+_{i\alpha\sigma}\hat{c}^+_{i\beta\sigma'}\hat{c}_{i\alpha\sigma'}\hat{c}_{i\beta\sigma} + (\hat{c}^+_{i\alpha\uparrow}\hat{c}^+_{i\alpha\downarrow}\hat{c}_{i\beta\downarrow}\hat{c}_{i\beta\uparrow} + \hat{c}^+_{i\beta\uparrow}\hat{c}^+_{i\beta\downarrow}\hat{c}_{i\alpha\downarrow}\hat{c}_{i\alpha\uparrow})]\} \quad (17)$$

Here $\hat{c}_{\mathbf{k}\alpha\sigma}$ annihilates a spin $\sigma$ electron in orbital $\alpha$ with momentum $\mathbf{k}$ and $i$ labels the sites of a square lattice.

We used Kuroki parameters in constructing $K_{\alpha\beta}(\mathbf{k})$ [15]. For interaction parameters, we have used the values in the literature [14]. The value of Fermi level corresponding to $x=0.1$ is $\mu=10.936\,eV$ and temperature is set as $k_BT=0.25\,meV$. Two of these five bands are thrown away since those are far away from the Fermi surface. In this case, Fermi surfaces in three bands are split into five pockets. The BZs of these three bands have been divided into 16 patches.(fig. 1) $LaFeAsO$ has $D_{4h}$ symmetry. Using this symmetry we can generate all renormalized vertex functions from those in the irreducible region of BZ, which can reduce the calculation time to 1/8.

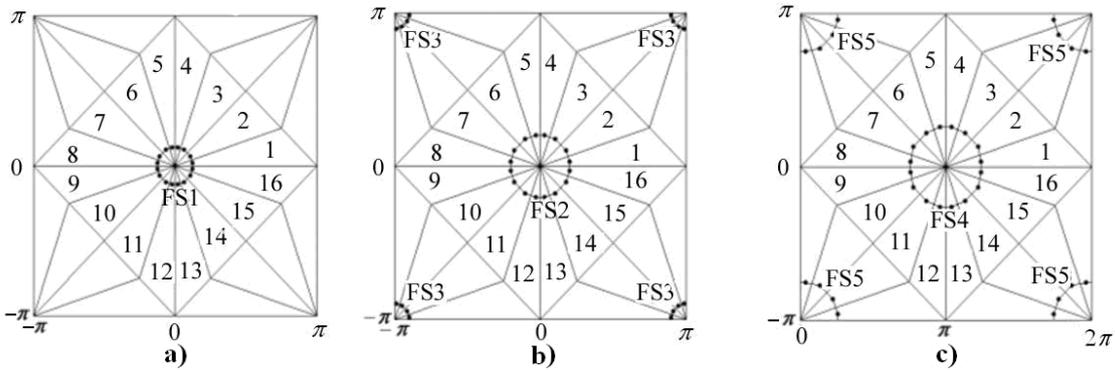

Fig. 1. A division of the BZ and the Fermi pockets.
a), b), and c) correspond to 1, 2, and 3-rd bands respectively.

Fig. 2 shows the RG flows of the strongly divergent susceptibilities of various instabilities at the doping $x=0.0\sim0.1$ as a function of the cut-off parameter $\Lambda$ of FRG. The most divergent susceptibilities in this doping range are those of $s$-wave pairing ($s$ SC) and spin density wave with $\mathbf{Q}=(\pi,0)$ ($SDW(\pi,0)$). They overwhelm the others and constitute main part of the phase diagram in this range. With increasing doping, the stripe AFM (SDW) order is quickly replaced by $s$-wave superconductivity, which is in good agreement with experiments and phase diagrams of the iron pnictides[5~7, 17].

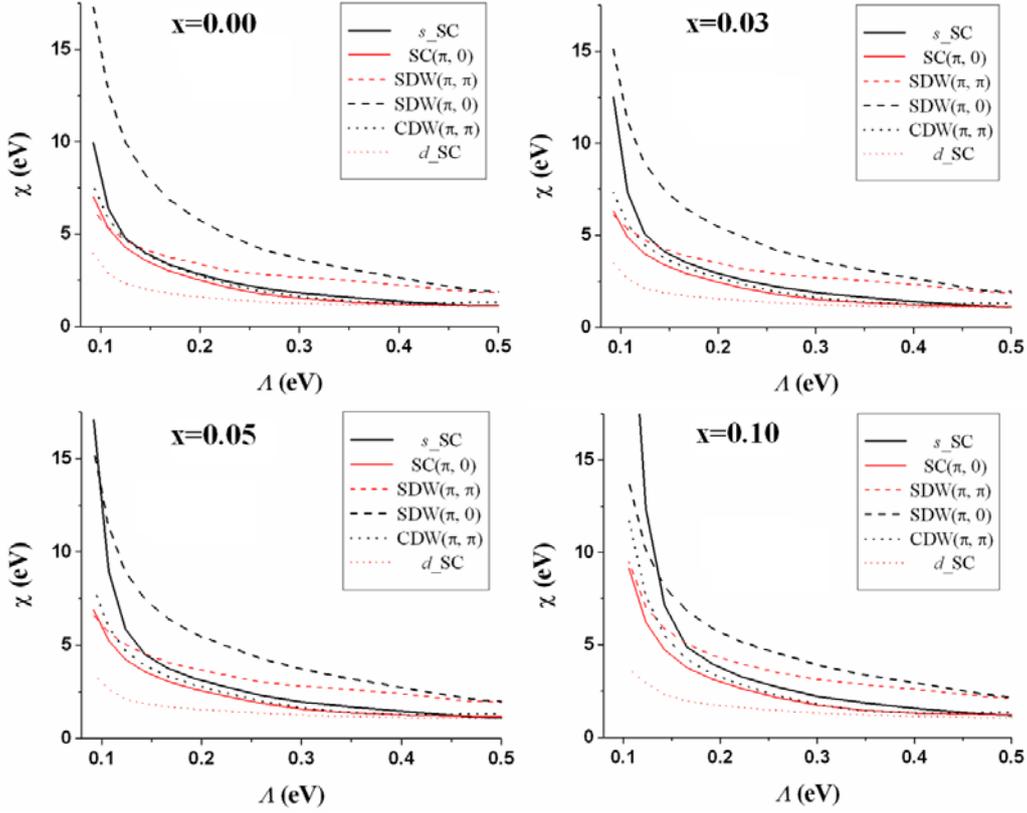

Fig. 2. The RG flows of diverse susceptibilities at the doping x=0.0~0.1 in LaFeAsO$_{1-x}$F$_x$.

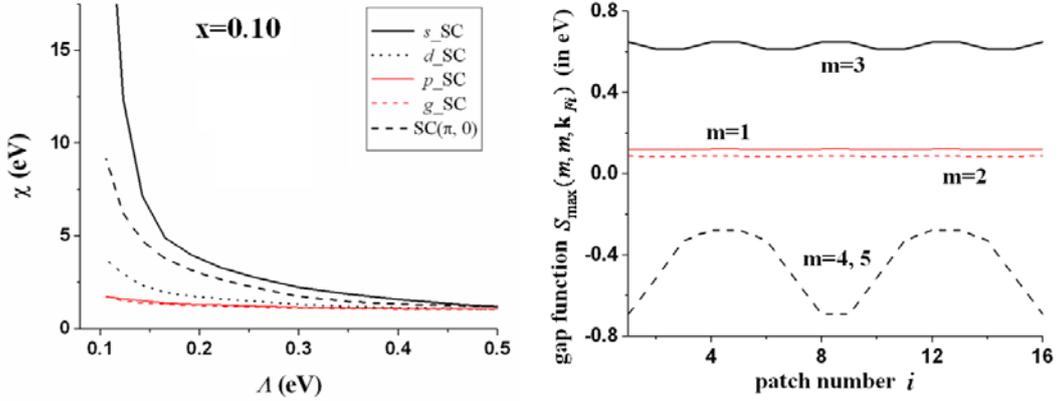

Fig. 3. The RG flows of the pairing susceptibilities in LaFeAsO$_{1-x}$F$_x$ (x=0.1)

Fig. 4. The superconducting gap function $S_{max}(m,m,\mathbf{k}_{Fi})$

The RG flows of the strongly divergent pairing susceptibilities have been shown in Fig. 3. As you see, a *s*-wave pairing susceptibility is overwhelmingly divergent, which is coincident with a result in the literature [14]. Thus, we conclude that the symmetry of the gap function in this superconductor (it agrees with the paring symmetry.) is an extended *s*-wave. It is noticeable that the staggered superconductivity has the secondary largest susceptibility. This means that the staggered superconducting phase with $\mathbf{Q}=(\pi,0)$ is expected to be the second favorable superconducting phase, which has not been suggested in previous works. The third largest susceptibility is that of *d*-wave pairing as in the cuprates.

The *s*-wave gap function $S_{max}(m,m,\mathbf{k}_{Fi})$ has been shown in Fig. 4. It is remarkable that the gap function takes an opposite sign on the electron ($m=4,5$) and hole ($m=1,2,3$) pockets, which is consistent with conclusions in many theoretical[14~16] and experimental[17~19] studies. This feature of opposite sign has been also obtained for several Fe based superconductors including LaFePO[20]. But our result is different from that of [14] in the fact that $S_{max}(m,m,\mathbf{k}_{Fi})$ has dominant values on the pockets $m=3,4,5$, not on $m=1,4,5$.

## 4. Conclusion

We have proposed an approach for determining various kinds of susceptibilities and order parameters in multi-band Hubbard model within patch approximation of FRG method. This approach has an advantage that it needs only the final FRG result of vertex function.

We applied it to the iron pnictide $LaFeAsO_{1-x}F_x$ to study diverse instabilities and evaluate the paring symmetry of this superconductor. We have explained theoretically the experimental fact that the stripe AFM order is quickly replaced by the superconductivity when increasing the doping. We conclude that the paring symmetry of this material is *s*-wave with the opposite sign on the electron and hole pockets. We also assert that the staggered superconducting phase with $\mathbf{Q} = (\pi, 0)$ has the possibility to occur.

## References

[1] Y. Kamihara et al.; J. Am. Chem. Soc., 130, 11, 3296, 2008.
[2] J. Zhao et al.; Nature Mateirals, 7, 953, 2008.
[3] J. Yang et al.; Supercond. Sci. Technol., 21, 082001, 2008.
[4] K. Haule et al.; Phys. Rev. Lett., 100, 22, 226402, 2008.
[5] C. Cruz et al.; Nature, 453, 899, 2008.
[6] K. Ishida et al.; J. Phys. Soc. Jpn., 78, 062001, 2009.
[7] R. H. Liu et al.; Phys. Rev. Lett., 101, 8, 087001, 2008.
[8] S. L. Skornyakov et al.; Phys. Rev., B.80, 9, 092501, 2009.
[9] J. Q. Yan et al.; Phys. Rev., B.78, 2, 024516, 2008.
[10] C. Cao et al.; Phys. Rev., B.77, 22, 220506(R), 2008.
[11] H. W. Ou et al.; arXiv, 0803.4328, 2008.
[12] R. Shankar; Rev. Mod. Phys., 66, 1, 129, 1994.
[13] C. Honerkamp et al.; Phys. Rev., B.63, 3, 035109, 2001.
[14] F. Wang et al.; Phys. Rev. Lett., 102, 4, 047005, 2009.
[15] K. Kuroki et al.; Phys. Rev. Lett., 101, 8, 087004, 2008.
[16] I. I. Mazin et al.; Phys. Rev. Lett. 101, 5, 057003, 2008.
[17] Y. Qiu et al.; Phys. Rev., B.78, 5, 052508, 2008.
[18] T. Y. Chen et al.; Nature, 453, 1224, 2008.
[19] K. Hashimoto et al.; Phys. Rev. Lett., 102, 20, 207001, 2009.
[20] F. Wang et al.; arXiv, 1002.3358, 2010.